\documentclass[
 aip,
 jmp,
 amsmath,amssymb,
 reprint,
]{revtex4-1}

\usepackage{graphicx}% Include figure files
\usepackage{dcolumn}% Align table columns on decimal point
\usepackage{bm}% bold math
\begin{document}

\preprint{AIP/123-QED}
%---------------------------------------- Title and Authors ----------------------------------------
%===================================================================================================
\title%[Sample title]%
{Ferromagnetic resonance and interlayer exchange coupling in magnetic multilayers with compositional gradients}
 
\author{D. M. Polishchuk}
\email{dpol@kth.se.}
\author{A. F. Kravets}
\author{Yu.~O.~Tykhonenko-Polishchuk}
\affiliation{Nanostructure Physics, Royal Institute of Technology, Stockholm, Sweden}
\affiliation{Institute of Magnetism, NAS of Ukraine, Kyiv, Ukraine}
 
\author{A.~I.~Tovstolytkin}
\affiliation{Institute of Magnetism, NAS of Ukraine, Kyiv, Ukraine}
 
\author{V. Korenivski}%
\affiliation{Nanostructure Physics, Royal Institute of Technology, Stockholm, Sweden}%
 
\date{\today}
%===================================================================================================

\begin{abstract}

Ferromagnetic resonance (FMR) in magnetic multilayers of type F1/f/F2, where two strongly ferromagnetic layers F1 and F2 are separated by a weakly magnetic spacer f with a compositional gradient along its thickness, is investigated. The method allows to detect the weak signal from the spacer in additional to the more pronounced and readily measured signal from the outer strongly-magnetic layers, and thereby study the properties of the spacer as well as the interlayer exchange interaction it mediates. Variable temperature FMR measurements, especially near the relevant Curie points, reveal a rich set of properties of the exchange interactions in the system. The obtained results are useful for designing and optimizing nanostructures with thermally-controlled magnetic properties. 

\end{abstract}


\maketitle

%---------------------------------------- I - Introduction -----------------------------------------
%===================================================================================================

\section{\label{sec:level1}Introduction}
 
Magnetic multilayer structures of the spin-valve type F1/NM/F2, where two ferromagnetic (FM) layers F1 and F2 are separated by a nonmagnetic (NM) spacer, exhibit rich physics including spin-dependent scattering,\cite{r1,r2} interlayer exchange coupling,\cite{r3,r4} spin-torque effects,\cite{r5,r6} etc. Due to this functional versatility, spin-valve type nanostructures have become the foundation for many spintronic applications.\cite{r7,r8,r9} Further technological progress will be based on the ability to further expand the functionality, with the use of external electric field, temperature, etc., to control the magnetic state in the nanostructure.
 
Recent studies have demonstrated that modified F1/f/F2 systems, where the weakly FM spacer (f) has a Curie temperature ($T_\text{C}^\text{f}$) much lower than that of the strongly FM outer layers (F1 and F2), can expand the functionality of the spin-valve family, yielding nanostructures with thermally-controlled magnetic properties.\cite{r10,r11} Depending on whether spacer f is in FM or PM (paramagnetic) state, two strongly FM layers can be exchange coupled (for $T$ $<$ $T_\text{C}^\text{f}$) or uncoupled (for $T$ $>$ $T_\text{C}^\text{f}$). For such nanostructure placed in an appropriately chosen magnetic field, temperature variation may lead to switching between parallel and antiparallel mutual alignment of the magnetic moments of the F1 and F2 layers. Thus, a variation in temperature and/or field can produce switching between the P and AP configurations of the nanostructure (the so-called Curie switch, CS, or Curie valve action), which can be used in spin-switch sensors, oscillators, and memory elements with intrinsic thermoelectronic control.\cite{r10,r12}
 
The experiments described in Ref.~\onlinecite{r13,r14} confirmed the concept of parallel-antiparallel switching in F1/f/F2 nanostructures, in particular incorporating Ni$_x$Cu$_{100-x}$ spacers enclosed by exchange-pinned Co$_{90}$Fe$_{10}$ and free Ni$_{80}$Fe$_{20}$ (Py) layers. A serious disadvantage of such structures is a relatively wide temperature range of the transitional state between the P and AP configurations. This mainly occurs due to the high nonuniformity of the effective interatomic exchange coupling and the magnetization distribution throughout the spacer thickness, especially in the proximity to the strongly ferromagnetic interfaces. The use of a gradient-type spacer f* = PM/weak FM/PM can greatly improve the effective-exchange uniformity and lead to a significantly narrower magnetic transition range, as was demonstrated in Ref.~\onlinecite{r11}. Unfortunately, traditional measurement tools, such as magnetometry, are unable to probe and characterize the state of the f* spacer, which impedes the development of competitive CSs with reliably predictable properties. Ferromagnetic resonance (FMR), which was particularly useful for characterization of CS structures with uniform spacers,\cite{r14} should be expected to also detect the key magnetic characteristics of specifically f* spacers, as well as the interlayer exchange properties they mediate.
 
In this work we study FMR in CS-multilayers containing gradient spacers. We directly detect the magnetic response from the spacer and conduct variable temperature studies of the interlayer exchange in the system near the spacer's Curie point.

%------------------------------- II - Samples and Experimental details -----------------------------
%===================================================================================================

\section{Samples and measurements}
 
\begin{figure*}
\includegraphics{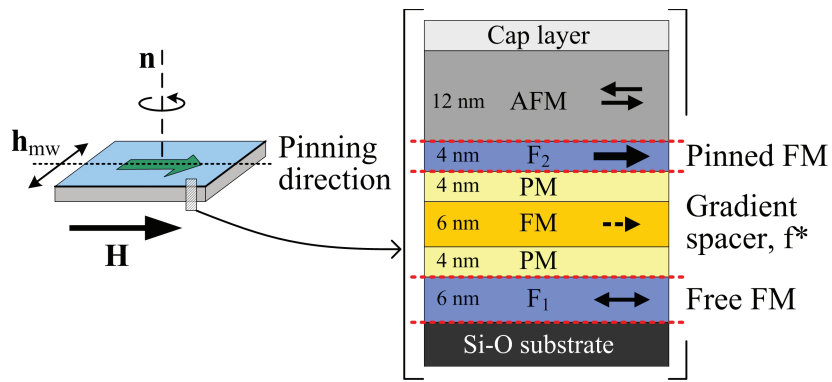}
\caption{Configuration used in magnetic resonance measurements (left panel) and layout of multilayers studied (right panel).}
\label{fig_1}
\end{figure*}

The configuration used in magnetic resonance measurements and the layout of the multilayer system F1/f/F2/AFM, where AFM is the pinning antiferromagnetic layer, are illustrated in Fig.~\ref{fig_1}. Thin-film magnetic multilayers Py(6~nm)/f(14~nm)/Py(2~nm)/ Co$_{90}$Fe$_{10}$(2~nm)/Ir$_{20}$Mn$_{80}$(12~nm) (hereafter -- F1/f/F2/AFM system) were deposited at room temperature on thermally oxidized silicon substrates using magnetron sputtering in an AJA Orion 8-target system (details of the fabrication are described in Ref.~\onlinecite{r11,r13}). The exchange pinning between the Py(2 nm)/CoFe(2 nm) bilayer and Ir$_{20}$Mn$_{80}$(12 nm) was set in during the deposition of the multilayers using an in-plane magnetic field $H_{\text{dep}} \approx$ 1 kOe. Three samples in the series studied here, labelled S-gr, S-fm and S-pm, have different spacer compositions. The S-gr sample has a gradient-type PM/weak FM/PM spacer: f* = Ni$_{50}$Cu$_{50}$(4 nm)/Ni$_{72}$Cu$_{28}$(6 nm)/Ni$_{50}$Cu$_{50}$(4 nm). The S-fm and S-pm samples have uniform spacers f = Ni$_{72}$Cu$_{28}$(14 nm) and f = Ni$_{56}$Cu$_{44}$(14 nm), respectively. Diluted alloys Ni$_{72}$Cu$_{28}$ and Ni$_{56}$Cu$_{44}$ in the bulk form are, respectively, ferromagnetic and paramagnetic in the temperature interval of measurements used in this work, $T$ = 120--380~K.\cite{r11}
 
%##################### Figure 1 #####################
%===================================================
%===================================================
 
The FMR measurements were performed using an X-band Bruker ELEXSYS E500 spectrometer equipped with an automatic goniometer. The operating frequency was $f$ = 9.46 GHz. In-plane FMR spectra were obtained with the external magnetic field applied $H$ parallel (P) and antiparallel (AP) to the AFM pinning direction (Fig.~\ref{fig_1}) at $T$ = 120--380~K.

%------------------------------ III - Results and Discussion -----------------------------
%===================================================================================================

\section{Results and Discussion}
 
\subsection{Experimental FMR spectra}
 
\begin{figure*}
\includegraphics{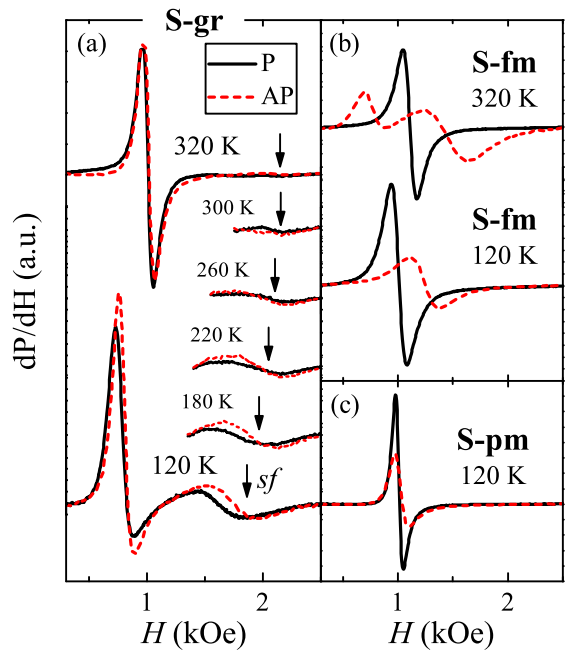}
\caption{Typical FMR spectra measured with the external magnetic field applied parallel (P) and antiparallel (AP) to the direction of the exchange pinning for (a) S-gs, (b) S-fm, and (c) S-pm samples. Panel (a) shows only the sf resonance signal for temperatures between 120 and 320 K.}
\label{fig_spectra}
\end{figure*}

Fig.~\ref{fig_spectra} shows typical FMR spectra measured at different temperatures for the P and AP orientations of field \textbf{H}. The signal of the highest intensity in all panels originates from the free layer, Py(6 nm). For the S-gr sample, at higher temperatures, the position of this signal is weakly dependent on the orientation of the magnetic field, while at lower temperatures the dependence becomes noticeable (Fig.~\ref{fig_spectra}(a)). In addition to the Py resonance line, the spectra for S-gr contain the line (labeled sf), with the resonance field close to 2 kOe, which is not observed in the spectra for the S-pm or S-fm samples. The intensity of the sf signal is highly dependent on temperature: it decreases with the increase in $T$ and vanishes as $T$ exceeds 320~K. The P and AP resonance fields of the sf signal are different at lower temperatures, but the lines approach each other and eventually merge as temperature increases. At the same time, their average position is shifted to higher fields compared to the signal from Py, which means that the effective magnetization of this component is lower than that of Py.\cite{r14} Since bulk Ni$_{72}$Cu$_{28}$ alloy has lower magnetization than Py, the sf signal can be ascribed to the Ni$_{72}$Cu$_{28}$(6 nm) layer in the gradient spacer PM/weak FM/PM. The Curie temperature of bulk Ni$_{72}$Cu$_{28}$ is relatively low ($\approx$ 330 K), which explains the strong temperature variation of the sf resonance field in the studied temperature region.
 
%##################### Figure 2 #####################
%===================================================
%===================================================
 
The highly intensive resonance line from the free layer F1 in the S-fm sample (Fig.~\ref{fig_spectra}(b)) has different resonance fields for the P and AP measurement configurations ($H_{\text{r}1}^\text{P}$ and $H_{\text{r}1}^\text{AP}$) in the whole temperature range. The nonzero difference ($H_{\text{r}1}^\text{AP} - H_{\text{r}1}^\text{P}$) indicates that the exchange pinning field from AFM, which acts on the F2 layer, “transmits” to the F1 layer due to a relatively strong interlayer coupling in the S-fm stack.\cite{r14} At higher temperatures, an additional low-field signal is observed, but only for the AP orientation of the magnetic field. This signal should be assigned to the pinned F2 layer, as detailed in Ref.~\onlinecite{r14}. The fact that this signal is almost indistinguishable in the AP spectra for the S-pm and S-gr samples needs additional investigation.
 
The same position of the P and AP resonance lines for the S-pm sample (Fig.~\ref{fig_spectra}(c)) means that the F1 and F2 layers are fully decoupled, which is the case at all measurement temperatures.

\subsection{Analysis of experimental data and discussion}
 
In our earlier FMR studies,\cite{r14,r15} a phenomenological theory was developed\cite{r16,r17,r18} for modelling the dynamics of CS-systems containing compositionally uniform spacers. Based on this theory, a comprehensive quantitative analysis of the experimental FMR results for CSs with spacers of different alloy dilution and thickness was performed, and the key magnetic parameters were determined. In this work, we will show that the same theoretical approach forms a reliable basis for understanding the temperature-dependent behavior of the resonance parameters for multilayers with gradient spacers (such as S-gr).

%##################### Figure 3 #####################
%===================================================
\begin{figure*}
\includegraphics{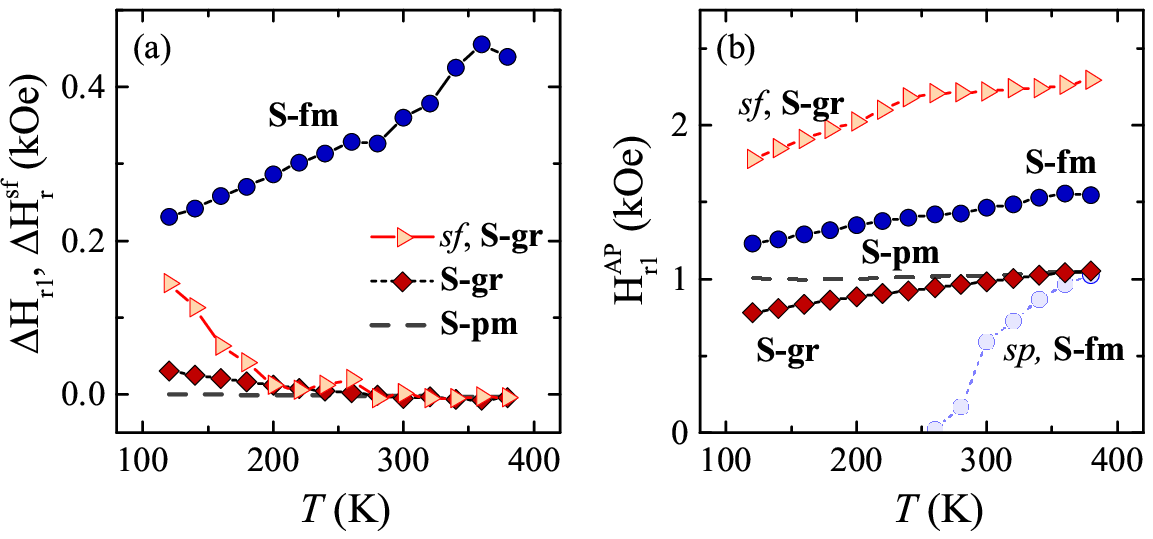}
\caption{(a) Resonance field difference $\Delta H_\text{r1} = H_{\text{r}1}^\text{AP}-H_{\text{r}1}^\text{P}$ versus temperature; $\Delta H_\text{r}^\text{sf}$ relates to the sf signal (see text for details). (b) Temperature dependences of the AP resonance field for all detected FMR signals. Shown additionally are the F1 resonance signals for all samples, the sf signal for the S-gr sample, and the F2 signal (labeled sp) for the S-fm sample.}
\label{fig_results}
\end{figure*}
%===================================================
 
Fig.~\ref{fig_results}(a) presents the temperature dependences of the resonance-field difference indicative of the exchange through the spacer, $\Delta H_{\text{r}1} = H_{\text{r}1}^\text{AP} - H_{\text{r}1}^\text{P}$, where $H_{\text{r}1}^\text{P}$, $H_{\text{r}1}^\text{AP}$ are the resonance fields for the F1 layer (in our case – Py) with the external field applied parallel and antiparallel to the AF pinning direction, respectively. According to eq.~(14) in Ref.~\onlinecite{r14}, $\Delta H_{\text{r}1} \sim \kappa^2 \cdot H_\text{b}$, where $\kappa$ is the constant of interlayer coupling, which depends on the magnetizations and thicknesses of the F1 and F2 layers and also includes the effective magnetization ($m$) and magnetic exchange length ($\Lambda$) of the weakly ferromagnetic spacer; $H_\text{b}$ is the effective biasing field acting on magnetization M2 of the F2 layer and is the measure of AFM pinning. Thus, for a constant $H_\text{b}$,  $\Delta H_{\text{r}1}$ reflects the strength of the interlayer coupling between strongly FM layers through the weakly FM spacer.
 
In our case, $\Delta H_{\text{r}1}$ is relatively high for S-fm (F1 and F2 are fully coupled) and zero for S-pm (F1 and F2 fully decoupled) in the whole temperature range, which is in agreement with the expected behavior. The decrease of both $\Delta H_{\text{r}1}$ and $\Delta H_\text{r}^\text{sf}$ for the S-gr sample with increasing temperature implies that the interlayer coupling, being significant at low temperatures, weakens as the  temperature increases.
 
The nontrivial effect that should be pointed out is the increase of $\Delta H_{\text{r}1}$ observed in the S-fm sample as temperature is increased (Fig.~\ref{fig_results}(a)). When the interlayer exchange coupling between F1 and F2 is weak, the temperature behavior of $\Delta H_{\text{r}1}$ is manly governed by $\kappa^2 \sim m^4$. $m$ decreases when temperature increases, which should lead to a decrease in $\Delta H_{\text{r}1}$. However, when the interlayer coupling is sufficiently strong, the assumption of a constant $H_\text{b}$ becomes invalid. In this case, the F1/f/F2 stack behaves like a single magnetic layer, and the effective biasing field from AFM acts on the whole F1/f/F2 structure, rather than only on F2. When the interlayer coupling between F1 and F2 becomes weaker, the action of the biasing field becomes localized to F2. As a result, the biasing effect becomes stronger and $\Delta H_{\text{r}1}$ increases. It is this situation that is believed to manifest in the S-fm sample as temperature is increased.
 
The resonance signal from the weakly FM spacer was undetectable in our earlier studies on CS-structures having spacers of uniform composition.\cite{r14} A possible reason for this is a strong influence of the adjacent FM layers on the magnetic state of the spacer. The gradient-type spacer studied herein suppresses the ferromagnetic proximity effect at the interface with the strongly ferromagnetic layers, and its core is therefore much more uniform magnetically, with presumably a much better defined FMR. This should make it easier to detect the signal from the inner part of the spacer (sf signal in Fig.~\ref{fig_spectra}(a)). At the same time, the nonzero difference $\Delta H_{\text{r}1}(T)$ for F1 at $T < 250$ K (Fig.~\ref{fig_results}(a)) indicates that the gradient-type spacer is still fully capable to mediate interlayer coupling at lower temperatures.
 
The value of $\Delta H_\text{r}^\text{sf}(T)$ is greater than $\Delta H_{\text{r}1}(T)$ at low temperatures ($T < 250$ K). This demonstrates that the effective transfer of the exchange pinning via the spacer can be considered to take place in two steps. First, the pinning is induced in the inner FM part of the spacer via the exchange coupling at the F2/f* interface. Then, the free F1 layer senses the pinning via the exchange coupling at the f*/F1 interface.
 
The temperature dependence of the AP resonance fields, shown in Fig.~\ref{fig_results}(b), contain additional information about the magnetic state of the studied CSs. Significant changes in $H_{\text{r}1}^\text{AP}$ versus temperature can be caused by two factors: (i) changes in the magnetization of FM, and (ii) changes in the strength of the interlayer coupling (due to changes in the magnetization of the spacer).
 
The resonance field of the free layer (F1) in the S-pm sample slightly increases with increasing temperature (Fig.~\ref{fig_results}(b)). In the FMR configuration used, an increase of $H_{\text{r}1}$ should correspond to a decrease in the F1 magnetization.\cite{r18} The $H_{\text{r}1}^0(T)$ dependence for S-pm is used as a reference temperature dependence.
%tut ja ne ponjal o chem i kak svjazano s prediduschim § /vk
 
$H_{\text{r}1}^\text{AP}(T)$ for the S-gr and S-fm samples differ from $H_{\text{r}1}^0(T)$ for S-pm. The observed changes here should be predominantly due to changes in the strength of the interlayer exchange. The strong interlayer coupling inherent to S-fm leads to a clear separation of $H_{\text{r}1}^\text{P}$ and $H_{\text{r}1}^\text{AP}$ from $H_{\text{r}1}^0(T)$. This is why $H_{\text{r}1}^\text{AP}(T)$ for S-fm is so high above that for S-pm. The weaker interlayer coupling inherent to S-gr affects the resonance field of F1 to a lesser extent. On the other hand, the $H_{\text{r,sf}}^\text{AP}(T)$ dependence of the sf signal from the spacer of S-gr can not be defined by only one dominant mechanism: strong changes with temperature of both the magnetization and the interlayer exchange coupling are at play.
 
Finally, the sp signal from the pinned F2 layer is distinguishable only in the AP spectra for S-fm at higher temperatures (Fig.~\ref{fig_spectra}(b) and Fig.~\ref{fig_results}(b)). Due to the strong AFM pinning, F2 must have negative resonance field for the P orientation, which is not observable.\cite{r14} It is interesting that this signal has a significantly reduced intensity in the spectra for the S-gr and S-pm samples.

%------------------------------ IV - Conclusions -----------------------------
%===================================================================================================

\section{Summary}
 
FMR spectroscopy and its theoretical analysis provide an excellent tool for describing the magnetism in multilayers of F1/f/F2 type, and determining the key magnetic parameters of the nanostructure as a whole and the Curie-spacer in particular. F1/f*/F2 multilayers with gradient-type spacers exhibit clear peaks due to the FMR in the spacer, which are used to characterize this critical layer in terms of its magnetization and the exchange coupling it mediates between the outer ferromagnetic layers.
 
\begin{acknowledgments}
Support from the Swedish Research Council (VR Grant No.~2014-4548), the Swedish Stiftelse Olle Engkvist Byggm\"astare, the Science and Technology Center in Ukraine (project P646), and the National Academy of Sciences of Ukraine (project 0115U003536 and 0115U00974) is gratefully acknowledged.
\end{acknowledgments}
 
%\nocite{*}
\bibliography{References}% Produces the bibliography via BibTeX.
 
\end{document}